\documentclass[twocolumn]{aastex6}
\usepackage{amssymb,amsfonts,amsmath,amstext,amsgen,amsopn,amsxtra,indentfirst,times}%

\usepackage{savesym}%
\usepackage{graphicx}%
\usepackage{hyperref}%
\usepackage{lineno}%
\usepackage{mathtools}%
\usepackage{amsmath}%
\usepackage{xspace}%
\usepackage{units}%
\usepackage{tikz}%
\usetikzlibrary{positioning}%
\usetikzlibrary{shapes,arrows}%
\usepackage{nicefrac}%
\usepackage{array}

\usepackage{tabularx}  
\usepackage{ragged2e}  
\newcolumntype{Y}{>{\RaggedRight\arraybackslash}X} 
\usepackage{booktabs}  

\newcommand{\rev}[1]{{\color{black} {#1}}}
\newcommand{\reve}[1]{{\color{black} {#1}}}
\newcommand{\renv}{$r_{\rm env}/{R}$\xspace}
\newcommand{\Zenv}{$Z_{\rm env}$\xspace}

\newcommand{\mice}{$m_{\rm water}$\xspace}

\newcommand{\RE}{R$_{\rm \Earth}$\xspace}
\newcommand{\ME}{M$_{\rm \Earth}$\xspace}

\newcommand{\Lenv}{$L_{\rm env}$\xspace}
\newcommand{\rc}{$r_{\rm core}$\xspace}
\newcommand{\rsolid}{$r_{\rm mantle}$\xspace}
\newcommand{\menv}{$m_{\rm env}$\xspace}

\newcommand{\fesi}{{\rm Fe}/{\rm Si}_{\rm bulk}\xspace}
\newcommand{\mgsi}{{\rm Mg}/{\rm Si}_{\rm bulk}\xspace}
\newcommand{\fesistar}{{\rm Fe}/{\rm Si}_{\rm star}\xspace}
\newcommand{\mgsistar}{{\rm Mg}/{\rm Si}_{\rm star}\xspace}
\newcommand{\fesima}{{\rm Fe}/{\rm Si}_{\rm mantle}\xspace}
\newcommand{\mgsima}{{\rm Mg}/{\rm Si}_{\rm mantle}\xspace}

\begin{document}

\title{Secondary atmospheres on HD~219134~b and~c}
\author{Caroline Dorn\altaffilmark{1}}
\author{Kevin Heng\altaffilmark{2,3,4}}

\altaffiltext{1}{University of Zurich, Institut of Computational Sciences, University of Zurich, Winterthurerstrasse 190, CH-8057, Zurich, Switzerland.  Emails: cdorn@physik.uzh.ch}
\altaffiltext{2}{University of Bern, Center for Space and Habitability, Gesellschaftsstrasse 6, CH-3012, Bern, Switzerland.  Emails: kevin.heng@csh.unibe.ch}
\altaffiltext{3}{Johns Hopkins University, Department of Earth and Planetary Sciences, 301 Olin Hall, Baltimore, MD 21218, U.S.A.}
\altaffiltext{4}{Johns Hopkins University, Department of Physics and Astronomy, Bloomberg Center for Physics and Astronomy, Baltimore, MD 21218, U.S.A.}

\begin{abstract}
   We analyze the interiors of  HD~219134~b and c, which are among the coolest super Earths detected thus far.  Without using spectroscopic measurements, we aim at \reve{constraining} if the possible atmospheres are \reve{hydrogen-rich or hydrogen-poor}.
   In a first step, we employ a full probabilistic Bayesian inference analysis in order to rigorously quantify the degeneracy of interior parameters given the data of mass, radius, refractory element abundances, semi-major axes, and stellar irradiation. We obtain constraints on structure and composition for core, mantle, ice layer, and atmosphere. In a second step, we aim to  draw conclusions on the nature of possible atmospheres by considering atmospheric escape. Specifically, we compare the actual possible atmospheres to a threshold thickness above which a \reve{primordial} (H$_2$-dominated) atmosphere  can be retained against evaporation over the planet's lifetime.
   The best constrained parameters are the individual layer thicknesses.  The maximum radius fraction of possible atmospheres are 0.18 and 0.13 $R$ (radius), for planets b and c, respectively. These values are significantly smaller than the threshold thicknesses of \reve{primordial} atmospheres: 0.28 and 0.19 $R$, respectively. Thus, the possible atmospheres of planets b and c are unlikely to be H$_2$-dominated. However, whether possible volatile layers are made of gas or liquid/solid water cannot be uniquely determined.
   Our main conclusions are: (1) the possible atmospheres for planets b and c are \reve{enriched and thus possibly secondary in nature}, and (2) both planets may contain a gas layer, whereas the layer of HD~219134~b must be larger. HD~219134~c can be rocky.

\end{abstract}


\section{Introduction}
\sloppy

\subsection{Motivation}

 Little is known about  compositional and structural diversity of super-Earths.
We often consider super-Earths to be distinct from sub-Neptunes in terms of their  {volatile fraction}. In fact, there is an intriguing transition around 1.5 \RE, above which most planets appear to contain a significant amount of volatiles \citep[e.g.,][]{ Rogers, Dressing}. The distribution of planet densities and radii suggest a transition that is continuous rather than stepwise \citep{leconte}, although the limited number of available observations might not allow a firm conclusion yet \citep{Rogers}.

A key criterion to distinguish super-Earths from sub-Neptunes is the origin of its atmosphere. Super-Earths atmospheres are thought to be dominated by outgassing from the interior, whereas sub-Neptunes have accreted and retained a substantial amount of \reve{primordial} hydrogen and helium. The atmospheric scale height will be significantly larger in the latter case since it scales as the reciprocal of the mean molecular mass. In consequence, the radius fraction of volatiles is often used to distinguish between  super-Earths and sub-Neptunes. 

The nature of an atmosphere, be it \reve{primordial} or secondary, helps to clearly categorize a planet. Atmospheres can have three different origins. (1) Accreted nebular gas from the protoplanetary disk (\reve{primordial} origin), (2) gas-release during disruption of accreting volatile-enriched planetesimals, or (3) outgassing from the interior (secondary origin). The time-scales associated with (1-2) and (3) are very different. An atmosphere that is dominated by outgassed planetesimal disruption (2) can theoretically be significantly different from a hydrogen-helium atmosphere \reve{\cite[e.g.,][]{fortney, elkins, schaefer, hashimoto, zahnle, venturini}}. \citet{venturini} show that enriched gas layers speed up the accretion of gas from the \reve{primordial} disk, which explains large fractions of H/He for intermediate mass planets. However, to what extent atmospheres of low-mass planets can be enriched (e.g., in water) and sustain their metallicity over their lifetime is subject of ongoing research.  For the close-in super-Earths HD~219134~b and c, we consider the two scenarios (1) and (3). \reve{In other words, we use the term {\it primordial} to refer to H$_2$-dominated atmospheres that are pristine and compositionally unaffected by subsequent physical or chemical processing including atmospheric escape \citep[e.g.,][]{Hu2015, Lammer} or interaction with the rocky interior. }

\begin{figure}[ht]
\centering
 \includegraphics[width = .5\textwidth, trim = 0cm 1cm 0cm 0cm, clip]{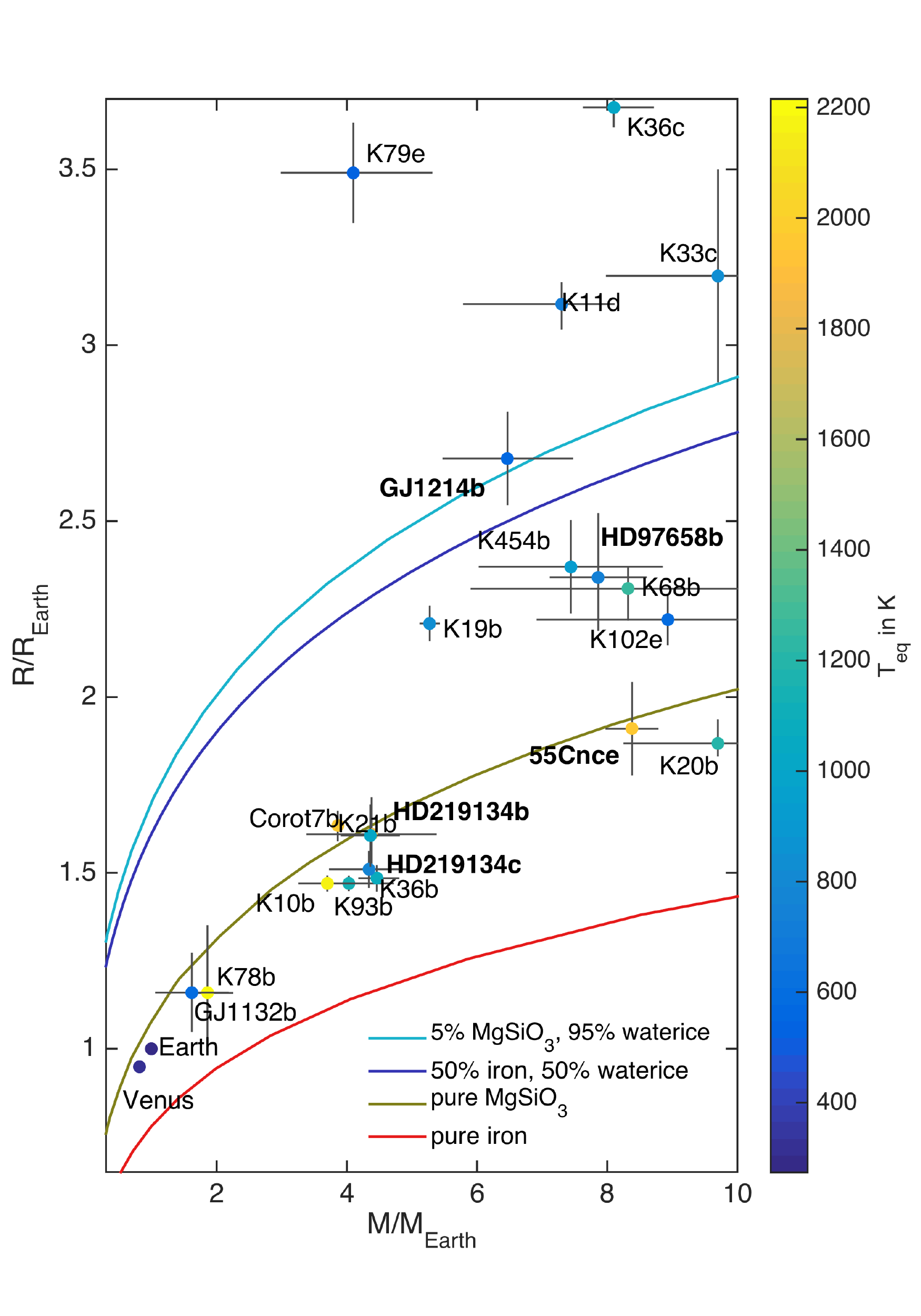}\\
 \caption{Mass-radius diagram for planets below 2.7 \RE and 10 \ME and mass uncertainties better than 20\% in general. HD~219134~b and c are among the coolest exoplanets yet detected regarding their equilibrium temperature (in color). Planets in bold face are included in the comparative study in section \ref{comparison}.}
 \label{fig1}
\end{figure}

The atmospheres of close-in planets are subject to significant mass loss (atmospheric escape), driven by extreme ultraviolet and X-ray heating from their stars.   The goal of this study is to present a method for determining if a planet may host a gaseous layer, and if this gas layer is \reve{ hydrogen-dominated (primordial) or dominated by high mean molecular masses} (secondary). \reve{Our method is different and complementary to studies that use spectroscopic signatures to distinguish between hydrogen-rich and hydrogen-poor atmospheres \citep[e.g.,][]{millerricci}.} We focus on the HD 219134 system, which hosts multiple planets.  Two of which fall in the super-Earth regime. Both planets b and c are transiting \citep{vogt, gillon} and represent together the coolest super-Earth pair yet detected in a star system (Figure \ref{fig1}).

The characterization of two planets from the same system benefits from possible compositional correlations between them. We can expect a correlation in relative abundance of refractory elements \citep{sotin07}. Abundances measured in the photosphere of the host star can be used as proxies for the relative bulk abundances, namely Fe/Si and Mg/Si \citep{dorn}. Here, we use different photospheric measurements on HD~219134, compiled by \citet{hinkel}. 
These bulk abundance constraints in addition to mass, radius, and stellar irradiation are the data that we use to infer structure and composition of the planets.

A rigorous interior characterization that accounts for data and model uncertainty can be done sensibly using Bayesian inference analysis, for which we use the generalized method of \citet{dornA}.
\rev{The previous work of \citet{dorn} and \citet{dornA} showed that Bayesian inference analysis is a robust method for quantifying interior parameter degeneracy for a given (observed) exoplanet. While \citet{dorn} focussed on purely rocky planets, a generalized method for super-Earths and mini-Neptunes was developed by \citet{dornA} by including volatiles (liquid and high pressure ices, and gas layers). Inferred confidence regions of interior parameters are generally large, which emphasizes the need to utilise extra data that further informs one about a planet's composition and structure. Here, we investigate additional considerations on atmospheric escape to further constrain the nature of the atmosphere. 

Similarly to the previous works, we assume planets that are made of distinct layers, i.e. iron core, silicate mantle, water layer, and gas layer as illustrated in Figure \ref{figsketch}. The use of an inference analysis allows us to account for the degeneracy among the layer properties, i.e., core size, mantle size and composition, water mass fraction, gas mass fraction and metallicity, and intrinsic luminosity. In this study, we account for interior degeneracy and calculate robust confidence regions of atmospheric thicknesses ($r_{\rm env}$). These inferred thicknesses $r_{\rm env}$ are then compared to theoretically possible thicknesses of a H$_2$-dominated atmosphere. The theoretically possible range of a H$_2$-dominated atmospheres is restricted due to atmospheric escape, i.e., too thin H$_2$-dominated atmospheres cannot be retained over a planet's lifetime. This implies a threshold thickness  below which H$_2$-dominated atmospheres cannot be retained. Here, we present how this threshold thickness ($\Delta R$) can be estimated. The comparison between $\Delta R$ and $r_{\rm env}$  is a key aspect of our study and allows us to draw conclusions about the nature of possible planetary atmospheres.

}

\subsection{Concept and method}

 {We wish to first describe the method conceptually, before providing the technical details later in the paper.  Consider a planet orbiting close to its star, which emits ultraviolet and X-ray radiation.  Planet formation occurs on short time-scales ($\sim 10^6-10^8$ years) and is essentially instantaneous over the lifetime of a $\sim 1$--10 Gyr-old star.  Immediately after the planet has formed, it retains a hydrogen-dominated atmosphere, which is then continuously eroded until the present time.  We take the total time lapsed to be the age of the star ($t_\star$).

The total mass of lost \reve{primordial} atmosphere, $M_{\rm env, lost}(t)$, increases over time due to atmospheric escape. Over the lifetime $t_\star$, the total escaped mass is $M_{\rm env, lost}(t_\star)$, which we convert to a fraction of the planetary radius, $\Delta R/R$. Atmospheric escape can erode $\Delta R$ worth of atmosphere over the age of the star. 
\rev{Independent of $\Delta R$ and from our Bayesian inference analysis, we can estimate the possible range of atmospheric thicknesses} at the present time, $r_{\rm env}(t_\star)$. If $r_{\rm env}(t_\star) < \Delta R$, then the atmosphere is not H$_2$-dominated, because any H$_2$ atmosphere would have been eroded away. {Thus, $\Delta R$ may be visualized as being the threshold thickness above which a \reve{primordial} atmosphere can be retained against atmospheric escape over a time $t_\star$. The comparison between $r_{\rm env}(t_\star)$ and $\Delta R$ is a key aspect of this study.} }



The outline of this study is as follows: We first discuss the method of characterizing  planet interiors. We explain how we approximate the amount of \reve{primordial} atmosphere that may be lost due to stellar irradiation and how we relate this to a threshold thickness of a \reve{primordial} atmosphere.  Based on these estimates, we demonstrate how we infer the atmospheric origin. We show results for HD~219134~b and c, and compare them to 55~Cnc~e, HD~97658~b, and GJ~1214~b. In an attempt to get an idea of the distribution of \reve{enriched (secondary)} atmospheres, we apply the method to low-mass planets ($<10$~\ME). We finish with a discussion and conclusions.

Note that we use the terms \emph{atmosphere} and \emph{gas layer} synonymously. The atmosphere/gas layer model comprises a radiative layer on top of a convection-dominated envelope.

\section{Methodology}
\label{Methodology}

\subsection{Interior characterization}
\begin{figure}[ht]
\centering
 \includegraphics[width = .4\textwidth, trim = 0cm 0cm 0cm 0cm, clip]{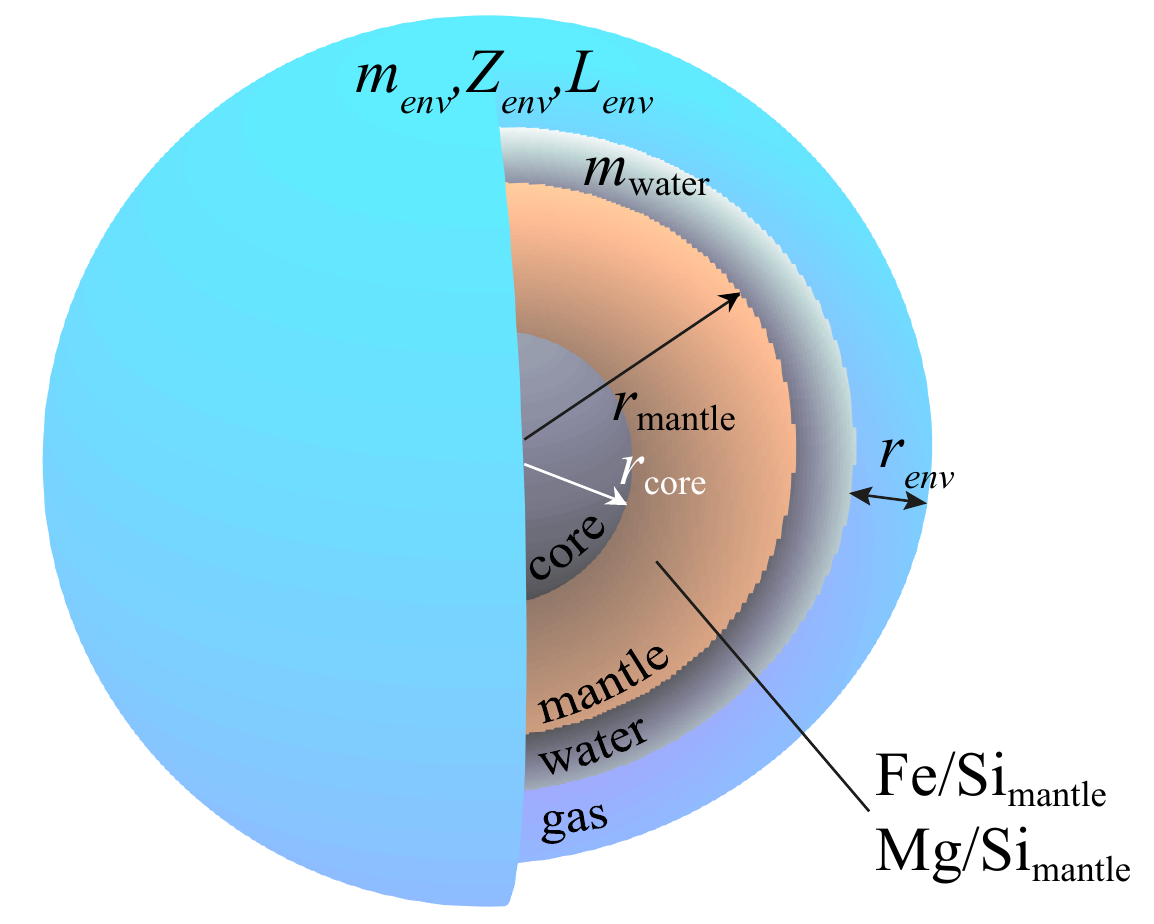}\\
 \caption{\rev{Illustration of interior parameters: core size (\rc), mantle composition ($\fesima$, $\mgsima$), mantle size (\rsolid), water mass fraction (\mice), intrinsic luminosity (\Lenv), gas mass (\menv), gas metallicity (\Zenv), and  atmospheric thickness ($r_{\rm env}$).}}
 \label{figsketch}
\end{figure}

 Using the generalized Bayesian inference analysis of \citet{dornA} \rev{that employs a Markov chain Monte Carlo (McMC) method}, we rigorously quantify the degeneracy of the following interior parameters for a general planet: 
\begin{itemize}
\item {\bf core}: core size (\rc),
\item  {\bf mantle}: mantle composition ($\fesima$, $\mgsima$) and size (\rsolid),
\item  {\bf water}: water mass fraction (\mice),
\item  {\bf gas}: intrinsic luminosity (\Lenv), gas mass (\menv), and metallicity (\Zenv). 
\end{itemize}
\reve{ From the posterior distribution of those interior parameters, we can compute the posterior distribution of the thickness of a possible gas layer ($r_{\rm env}$), which we then use to infer if the gas layer is hydrogen-rich or poor (Section \ref{compa}). Regarding the volatile-rich layers, our parameterization allows us to produce planet structures that range from (1) purely-rocky to (2) thick water layers with no additional gas layer to (3) thick gas layers without water layers below. The latter structure (3) determines the largest values of $r_{\rm env}$. }
Figure \ref{figsketch} illustrates the interior parameters of interest.

  The considered data comprise:
  \begin{itemize}
\item mass $M$,
\item radius $R$,
\item bulk abundance constraints on $\fesi$ and $\mgsi$, and minor elements Na, Ca, Al,
\item semi-major axes $a$,
\item stellar irradiation (namely, effective temperature $T_{\rm eff}$ and stellar radius $R_\star$).
\end{itemize}

For $\fesi$, $\mgsi$ and minor elements, we use their equivalent stellar ratios as proxies that can be measured in the stellar photosphere \citep{dorn}. 

The prior distributions of the interior parameters are listed in Table \ref{tableprior}. The priors are chosen conservatively. The cubic uniform priors on \rc and \rsolid reflect equal weighing of masses for both core and mantle. Prior bounds on $\fesima$ and $\mgsima$ are determined by the host star's photospheric abundance proxies. Since iron is distributed between core and mantle, $\fesi$ only sets an upper bound on  $\fesima$. 
A log-uniform prior is set for \menv and \Lenv.  A uniform prior in \Zenv equally favors metal-poor and metal-rich atmospheres, which seems appropriate for secondary atmospheres. In Section \ref{HDstuff}, we investigate the effect of different priors on \Zenv.

 In this study, the \reve{planetary} interior is assumed to be composed of a pure iron core, a silicate mantle comprising the oxides Na$_2$O--CaO--FeO--MgO--Al$_2$O$_3$--SiO$_2$, pure water layer, and an atmosphere of H, He, C, and O. 
 
The structural model for the interior uses self-consistent thermodynamics for core, mantle, high-pressure ice, and water ocean, and to some extent also atmosphere. 
For the core density profile, we use the equation of state (EoS) fit of iron in the hcp (hexagonal close-packed) structure provided by \citet{bouchet} on {\it ab initio} molecular dynamics simulations. \reve{We assume a solid state iron, since the density increase due to solidification in the Earth's core is small (0.4 g/cm3, or 3\%) \citep{Dziewonski}.}
For the silicate mantle, we compute equilibrium mineralogy and density as a function of pressure, temperature, and bulk composition by minimizing Gibbs free energy \citep{connolly09}. For the water layers, we follow \citet{Vazan} using a quotidian equation of state (QEOS) and above 44.3 GPa, we use the tabulated EoS from \citet{seager2007} that is derived from DFT simulations. \reve{Depending on pressure and temperature, the water can be in solid, liquid or vapour phase.}
We assume an adiabatic temperature profile within core, mantle, and water layers.
The surface temperature of the water layer is set equal to the temperature of the bottom of the gas layer.


For the gas layer, we solve the equations of hydrostatic equilibrium, mass conservation, and energy transport. For the EoS of elemental compositions of H, He, C, and O, we employ the CEA (Chemical Equilibrium with Applications) package \citep{CEA}, which performs chemical equilibrium calculations for an arbitrary gaseous mixture,  including dissociation and ionization and assuming ideal gas behavior. The metallicity \Zenv is the mass fraction of C and O in the gas layer, which can range from 0 to 1.  
\reve{For the gas layer, we assume an irradiated layer on} top of a \reve{convective-dominated} envelope, 
for which we assume a semi-gray, analytic, global temperature averaged profile \citep{GUILLOT2010, Heng2014}. 
The boundary between the irradiated layer and the  \reve{underlying} envelope is defined where the optical depth in visible wavelength is $100 / \sqrt{3}$ \citep{JIN2014}. Within the envelope, the usual Schwarzschild criterion is used to distinguish between convective and radiative layers. 
The planet radius is defined where the chord optical depth becomes 0.56 \citep{Lecavelier08}. 

We refer to model I in \citet{dornA} for more
details on both the inference analysis and the structural model.

\begin{table}[ht]
\caption{Prior ranges.  \label{tableprior}}
\begin{center}
\begin{tabular}{lll}
\hline\noalign{\smallskip}
parameter & prior range & distribution  \\
\noalign{\smallskip}
\hline\noalign{\smallskip}
$r_{\rm core}$         & (0.01  -- 1) $r_{\rm mantle}$ &uniform in $r_{\rm core}^3$\\
$\fesima$           & 0 -- $\fesistar$&uniform\\
$\mgsima$         & $\mgsistar$ &Gaussian\\
$r_{\rm mantle}$   & (0.01 -- 1) $R$& uniform in $r_{\rm mantle}^3$\\
$m_{\rm water}$ & 0 -- 0.98 $M$& uniform\\
\menv            & 0 -- $m_{\rm env, max}$  &uniform in log-scale\\
\Lenv                & $10^{18} - 10^{23}$ erg/s&uniform in log-scale\\
\Zenv                & 0 -- 1&uniform\\
\hline
\end{tabular} 
\end{center}
\end{table}

\subsection{Estimating the threshold thickness $\Delta R$ of a \reve{primordial} atmosphere \rev{layer} considering atmospheric escape}

We approximate $\Delta R$ by the atmospheric \rev{layer} thickness that corresponds to the \rev{accumulated} mass of hydrogen that may be lost over the planet's lifetime \rev{($M_{\rm env,lost}$)}. Loss rates are determined by X-ray irradiation from the star and mass-loss efficiencies. \reve{Hydrostatic balance} is used to calculate the layer thickness $\Delta R$ corresponding to a \reve{primordial} atmosphere of mass $M_{\rm env,lost}$. The detailed calculation of $\Delta R$ involves several steps, that are discussed in the following.

 \rev{Let the layer thickness $\Delta R$ be the difference in radius attributed to a \reve{primordial} atmosphere. If we assume this layer to be in hydrostatic equilibrium, then this difference in radius is}
\begin{equation}\label{eq0}
\Delta R = H \ln{\left(\frac{P_{\rm b}}{P_{\rm t}}\right)},
\end{equation}
where $H$ is pressure scale height and $P_{\rm b}$ is the pressure at the bottom of the \rev{layer}, which we will derive in the next paragraphs.
$P_{\rm t}$ is \rev{the pressure at the top of the layer, corresponding to the} transit radius \citep{heng16},
\begin{equation}
P_{\rm t} \approx \frac{g}{\kappa} \sqrt{\frac{H}{2\pi R}}.
\end{equation}
If we assume a mean opacity of $\kappa = 0.1$ cm$^2$ g$^{-1}$ \citep{freedman14}, then for both the b and c planets we get $P_{\rm t} \approx 1$ mbar.

The pressure scale height $H$ is calculated assuming a hydrogen-dominated layer (mean molecular mass $\mu=2~$g/mol) and using the equilibrium temperature $T_{\rm eq}$,
\begin{equation}\label{eq3}
H = \frac{T_{\rm eq} R^{*}}{g_{\rm surf} \mu },
\end{equation}
where $g_{\rm surf}$ is surface gravity and $R^{*}$ is the universal gas constant (8.3144598 J mol$^{-1}$ K$^{-1}$). The estimates in \citet{heng16} suggest that the assumption of $T = T_{\rm eq}$ is reasonable. Values for $g_{\rm surf}$ and  $T_{\rm eq}$ are listed in Tables \ref{data1} and \ref{data2}.

\rev{The pressure at the bottom of the layer $P_{\rm b}$  corresponds to the accumulated mass of escaped hydrogen over the planet's lifetime $M_{\rm env,lost}$,}
\begin{equation}\label{eq2}
P_{\rm b} = \frac{ g M_{\rm env,lost}}{4 \pi R^2},
\end{equation}
which is simply a restatement of Newton's second law. 
  $M_{\rm env,lost}$ is approximated by atmospheric escape considerations. Dimensional analysis yields an expression for the atmospheric escape rate,
\begin{equation}\label{eq1}
\dot{M} = \frac{\pi \eta F_{\rm X} R^2}{E_g},
\end{equation}
where $F_{\rm X}$ is the X-ray flux of the star and $E_g = GM/R$ is the gravitational potential energy.  The evaporation efficiency, $\eta$, is the fraction of the input stellar energy that is converted to escaping outflow from the planet.  It is often assumed to be a constant, but it is more likely that its value varies with the age of the system \citep{ow13}. The evaporation efficiency $\eta$ has been studied by various authors \citep[e.g.,][]{Shematovich,salz}, who demonstrate that values between 0.01 and 0.2 are reasonable for our planet range of interest.  In other words, $\eta$ hides the complexity of atmospheric radiative transfer of X-ray photons \reve{as well as unknown quantities such as the planetary albedo}.

\begin{figure}
\begin{center}
\includegraphics[width=\columnwidth]{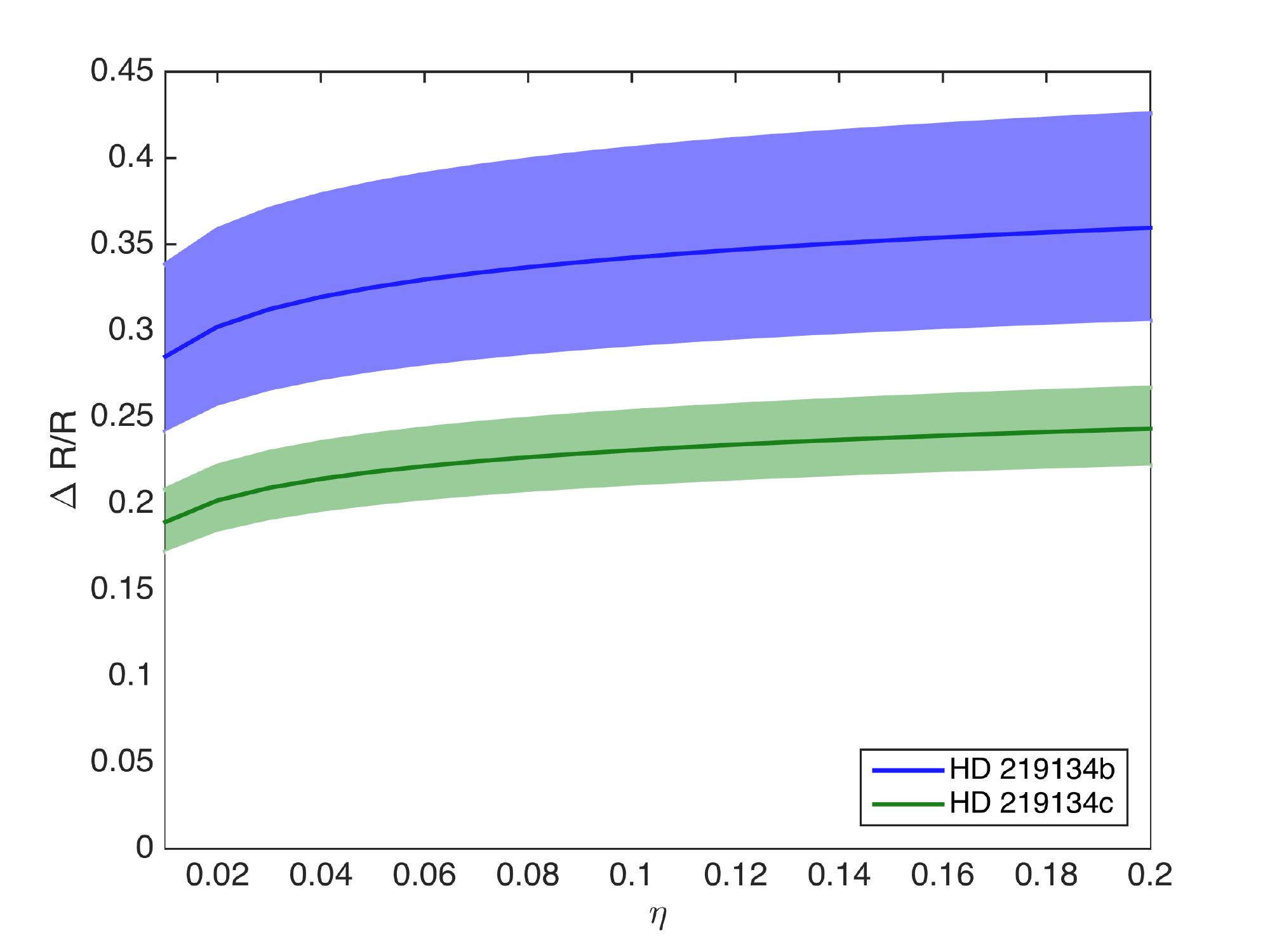}
\end{center}
\caption{Threshold thickness $\Delta R$ as a function of evaporation efficiency $\eta$. The spread accounts for the uncertainty in planet mass and radius, and age. If the inferred radius $r_{\rm env}$ is less than $\Delta R$, then the atmosphere is most likely enriched and not dominated by H$_2$.}
\label{fig:dr}
\end{figure}

The strongest assumption we make is that mass loss is constant over the planet's lifetime, such that $M_{\rm env,lost} =  t_\star \dot{M}$ ($t_\star=12.9$ Gyr; \citealt{takeda07}). 
 Thus, equations \ref{eq2} and \ref{eq1} provide us the expression
\begin{equation}
P_{\rm b} = \frac{\eta L_{\rm X} t_\star}{16 \pi a^2 R},
\end{equation}
with $L_{\rm X} = 4 \pi a^2 F_{\rm X} = 4 \times 10^{26}$ erg s$^{-1}$ \citep{porto06} being the X-ray luminosity of the star.  {In Figure \ref{fig:dr}, we compute $\Delta R/R$ as a function of $\eta$, since the exact value of $\eta$ is not well known.  Fortunately, $\Delta R/R$ depends weakly on $\eta$. Also, uncertainty in stellar age only has \reve{a small} effect on $\Delta R/R$: a difference in stellar age of 1 Gyr only introduces variations on $\Delta R/R$ of less than one percent. The spread in $\Delta R/R$ is mainly due to the uncertainties in planetary mass and radius. }

The physical interpretation of the preceding expressions for $P_{\rm b}$ and $\Delta R$ are worth emphasizing.  {The former is the amount of \reve{primordial} atmosphere that may be lost by atmospheric escape during the lifetime of the star.}  It provides a conservative estimate, because we have assumed the X-ray luminosity to be constant, whereas in reality stars tend to be brighter in X-rays earlier in their lifetimes:
\begin{equation}
M_{\rm env,lost} = \int_0^{t_\star} \dot{M}(t) dt > \dot{M}t_\star \,.
\end{equation}
The expression for $\Delta R$ is then a lower limit for a \rev{\reve{primordial}} atmosphere thickness corresponding to this atmospheric mass loss scenario.

\subsection{Assessing secondary/\reve{primordial} nature of an atmosphere}
\label{compa}
\begin{figure}[ht]
\centering
 \includegraphics[width = .5\textwidth, trim = 0cm 0cm 0cm 0cm, clip]{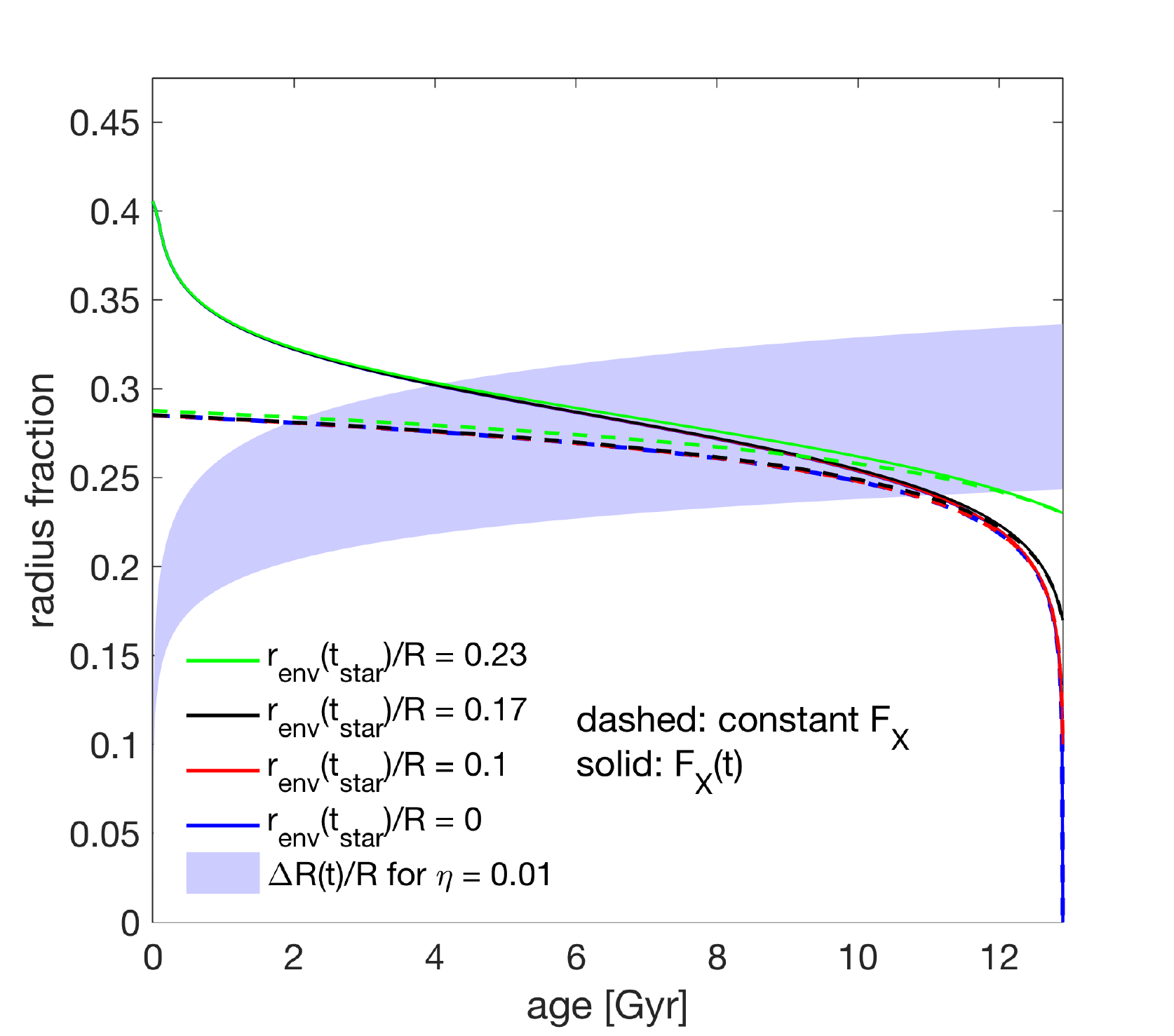}\\
 \caption{Evolution of H$_2$-dominated atmosphere thicknesses  \renv for HD219134b leading to different thicknesses at $t=t_\star$ ($\eta$ = 0.01 in all cases). Solid curves account for a time variable stellar X-ray flux $F_X(t)$  \citep{Ribas}, whereas dashed curves imply a constant $F_X$.The blue-shaded area depicts the evolution of $\Delta R/R$, its spread accounts for the uncertainties in planet mass and radius. }
 \label{fig9}
\end{figure}

We wish to compare $\Delta R$ with the gas thickness $r_{\rm env}$ inferred from the interior characterization.
There are three possible scenarios:
\begin{itemize}
\item $r_{\rm env} > \Delta R$:
Atmospheric escape is not efficient enough in removing \rev{a possible \reve{primordial}} atmosphere. This suggests that a large portion of the atmosphere can be \reve{primordial}. \rev{However, a secondary atmosphere is also possible.}

\item $r_{\rm env} \approx \Delta R $:
Mass loss can be still ongoing and no conclusion can be drawn about the nature of the atmosphere.

\item $r_{\rm env} <  \Delta R  $:
Atmospheric escape should have efficiently removed any \reve{primordial} H$_2$ atmosphere. If a finite \renv is inferred, the atmosphere is likely enriched and thus of secondary origin. Since the calculation of the threshold thickness $\Delta R$ is conservative, this  is the only scenario that can be used for a conclusive statement on the atmospheric origin.

This conclusion is illustrated in Figure \ref{fig9}, where the time-evolutions of H$_2$-dominated atmosphere thicknesses for HD219134~b are shown for $\eta=0.01$. The curves are constructed such that at $t=t_\star$ the relative thicknesses \renv are equal to 0 (blue), 0.1 (black), 0.17 (red), and 0.23 (green). Furthermore, the solid curves include the time-evolution of  X-ray flux, which we assume here to be solar-like ($F_{\rm X} \propto t^{-1.83}$ for $t > t_{\rm sat}$ and 
$F_{\rm X} = F_{\rm X}$ for $t < t_{\rm sat}$, where the saturation time is equal to 100 Myr and
 $F_{\rm X}(t_\star)$ is the observed value) \citep{Ribas}. Compared to a constant X-ray flux, the higher stellar activity for a young star implies that an atmosphere thickness of \renv at $t_\star$ must have started with a higher gas fraction at $t=0$ (see difference between solid and dashed curves in Figure \ref{fig9}). 
In both scenarios, we find that the smaller the observed thickness \renv compared to $\Delta R/R$,  the shorter the time a planet spends with this atmosphere thickness. 
\reve{Thus it is possible for a planet to host remaining small amounts of an initially thick \reve{primordial} atmosphere that will have a \renv lower than the threshold thickness. However, we find that this state is a very short fraction of the planet's lifetime, so it is unlikely that the planets we observe with \renv lower than the threshold value will be remnants of thicker \reve{primordial} atmospheres.}
In consequence, inferred atmospheres with thicknesses less than $\Delta R/R$ are likely to be \reve{enriched (secondary)}.

\end{itemize}

\section{Results}
\label{Results}

\subsection{Interiors of HD~219134~b and c}
\label{HDstuff}

\begin{figure*}[ht]
\centering
 \includegraphics[width = .8\textwidth, trim = 0cm 0cm 1.5cm 0cm, clip]{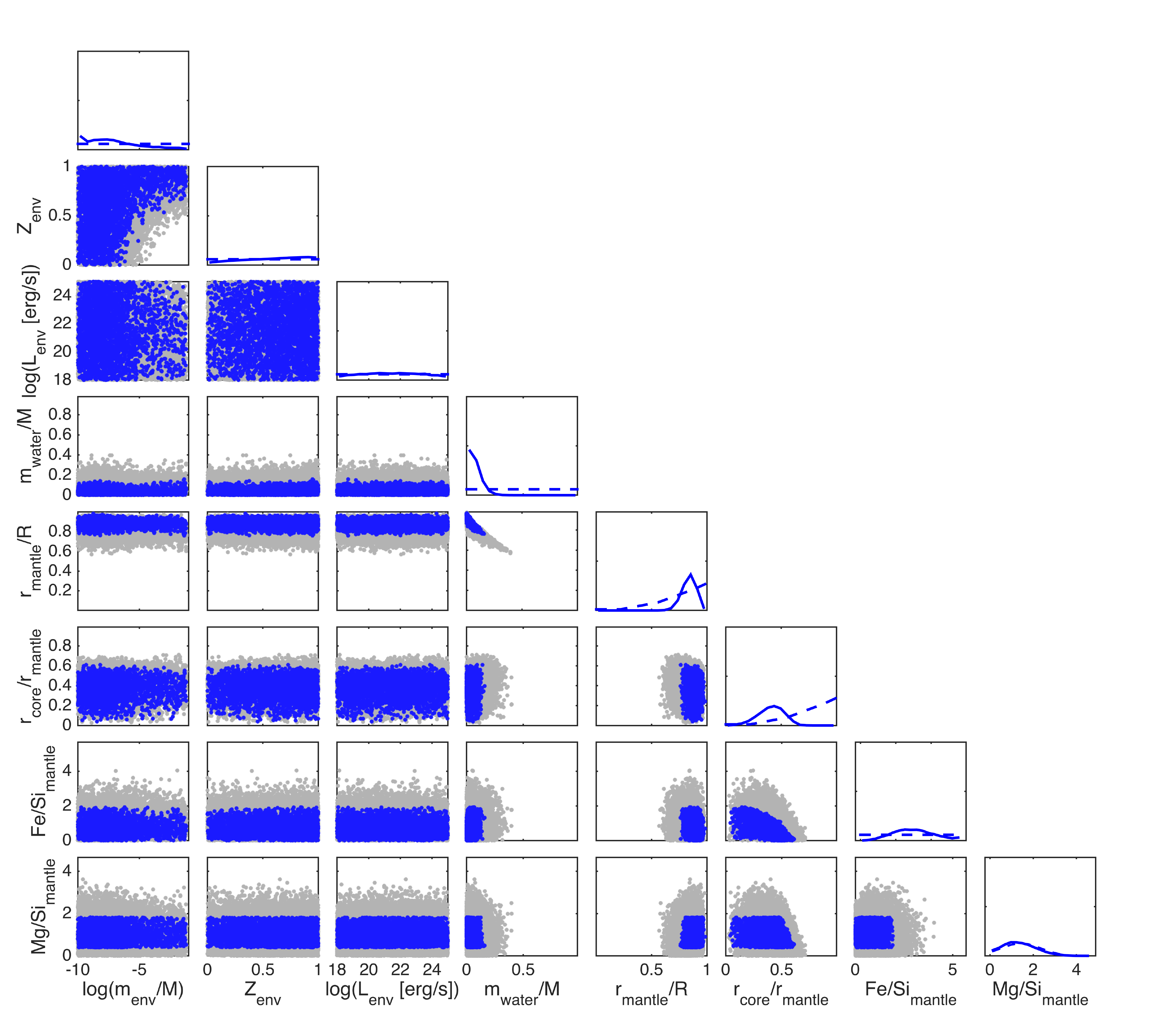}\\
 \caption{Sampled two and one-dimensional marginal posterior for HD~219134~b interior parameters: gas mass \menv, gas metallicity \Zenv, intrinsic luminosity \Lenv, mass of water \mice, radius of rocky interior \rsolid, core radius \rc, and mantle's relative abundances $\fesima$ and $\mgsima$. Blue dots explain the data within 1-$\sigma$ uncertainty. \reve{Dashed curves represent the prior distributions assumed.}}
 \label{corrB}
\end{figure*}

\begin{figure*}[ht]
\centering
 \includegraphics[width =.8\textwidth, trim = 0cm 0cm 1.5cm 0cm, clip]{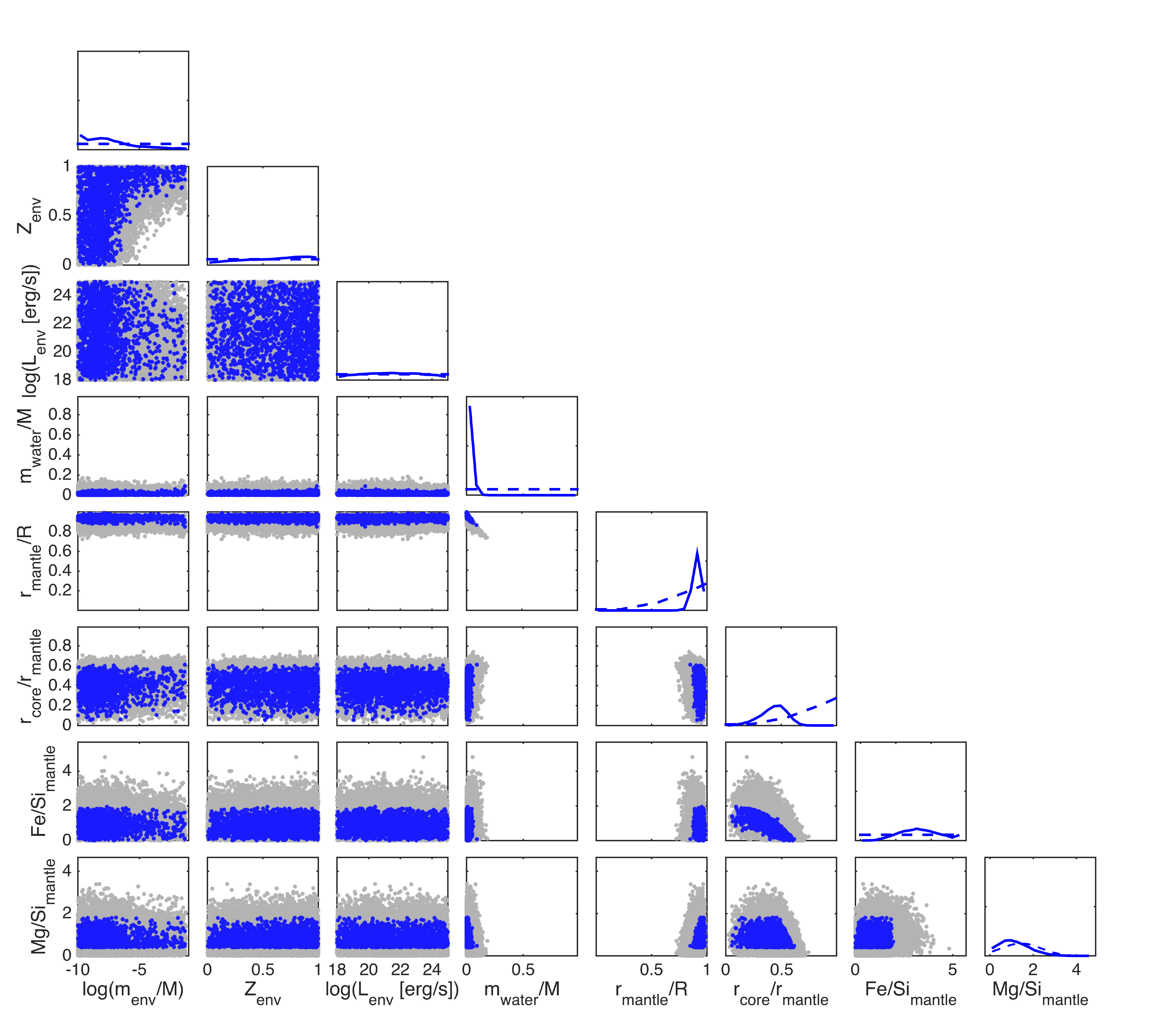}\\
 \caption{Sampled two-dimensional (2-D) marginal posterior for HD~219134~c interior parameters: gas mass \menv, gas metallicity \Zenv, intrinsic luminosity \Lenv, mass of water \mice, radius of rocky interior \rsolid, core radius \rc, and mantle's relative abundances $\fesima$ and $\mgsima$. Blue dots explain the data within 1-$\sigma$ uncertainty. \reve{Dashed curves represent the prior distributions assumed.}}
 \label{corrC}
\end{figure*}

We apply the inferrence method to HD~219134~b and c with the data listed in Tables \ref{data1} and \ref{data2}. The latter lists different stellar abundance estimates from the literature \citep[][]{Mishenina15,Ramirez,Valenti,Thevenin,Thevenin2} that were compiled by \citep{dornB} to \reve{examine} different bulk abundance scenarios. Besides a median abundance estimate (V0), they provide an iron-rich (V1) and an iron-poor (V2) scenario, that \reve{reflect} the limited accuracy in stellar abundance estimates \reve{(Table \ref{bulk})}. First, we use the median stellar abundance estimate denoted with V0.
Figures \ref{corrB} and \ref{corrC} show the two and one-dimensional (2-D and 1-D) marginal posteriors for all eight model parameters. 
Best constrained parameters are the layer thicknesses represented by \mice, \rsolid, \rc. We summarize our findings on the interiors of HD~219134~b and c with respect to the models that fit the data within 1-$\sigma$ uncertainty (blue dots in Figs. \ref{corrB} and \ref{corrC}):
\begin{itemize}
\item The possible interiors of HD~219134~b and c span a large region, including purely rocky and volatile-rich scenarios.
\item  {less than 0.1\%} of the model solutions for planets b and c, repectively,  are rocky ($r_{rocks}$/R $>$ 0.98).
\item The possible water mass fraction of HD~219134~b and c can reach from 0 -- 0.2 and 0 -- 0.1, respectively.
\item Unsurprisingly, the individual atmosphere properties (\menv, \Lenv, \Zenv) are weakly constrained. Consequently, their pdfs are dominated by prior information. However, the possible range of atmosphere thickness is well constrained to  0 -- 0.18 and 0 -- 0.13 for planets b and c, respectively (see Section \ref{vary}). 
\end{itemize}

\begin{table}[ht]
\caption{Summary of \reve{planetary} data \citep{motalebi, gillon}. \label{data1}}
\begin{center}
\begin{tabular}{lcccccc}
\hline\noalign{\smallskip}
parameter &\multicolumn{3}{c}{ HD~219134~b}& \multicolumn{3}{c}{HD~219134~c}  \\
\noalign{\smallskip}
\hline\noalign{\smallskip}
$R$/\RE &\multicolumn{3}{c}{ 1.606$\pm$0.086}&\multicolumn{3}{c}{1.515 $\pm$ 0.047}\\
$M$/\ME &\multicolumn{3}{c}{4.36$\pm$0.44}&\multicolumn{3}{c}{4.34 $\pm$ 0.22}\\
$g_{\rm surf}$[cm/$s^{-2}$] &\multicolumn{3}{c}{1656}&\multicolumn{3}{c}{1865}\\
$T_{\rm eq}$ [K] &\multicolumn{3}{c}{1025}&\multicolumn{3}{c}{784}\\
\vspace{2mm}
$a$ [AU] &\multicolumn{3}{c}{0.038}&\multicolumn{3}{c}{0.065}\\
\hline
\end{tabular} 
\end{center}
\end{table}

\begin{table}[ht]
\caption{Summary of \reve{stellar} data \citep{motalebi}. \label{data2}}
\begin{center}
\begin{tabular}{lc}
\hline\noalign{\smallskip}
parameter &\multicolumn{1}{c}{ HD~219134}\\
\noalign{\smallskip}
\hline\noalign{\smallskip}
$R_{\rm star}$/R$_{\rm sun}$ &0.778~$\pm$~0.005 \\
$T_{\rm eff}$ in K& 4699~$\pm$~16 \\
$[\rm Fe/H]$ & 0.04--0.84\\
\vspace{2mm}
$[\rm Fe/H]_{median}$ & 0.13   \\
$[\rm Mg/H]$ & 0.09--0.37 \\
\vspace{2mm}
$[\rm Mg/H]_{median}$ & 0.32  \\
$[\rm Si/H]$& 0.04--0.27\\
\vspace{2mm}
$[\rm Si/H]_{median}$ &0.12\\
$[\rm Na/H]$& 0.17--0.32 \\
\vspace{2mm}
$[\rm Na/H]_{median}$&0.19\\
$[\rm Al/H]$& 0.16--0.29 \\
\vspace{2mm}
$[\rm Al/H]_{median}$& 0.23\\
$[\rm Ca/H]$& 0.18--0.25\\
$[\rm Ca/H]_{median}$ & 0.21\\
\hline
\end{tabular} 
\end{center}
\end{table}

\begin{table}[ht]
\caption{Considered planet bulk abundance cases. V0 represents median abundance estimates, whereas V1 and V2 refer to iron-rich and iron-poor cases, respectively. \label{bulk}}
\begin{center}
\begin{tabular}{lccccccc}
\hline\noalign{\smallskip}
parameter& &\multicolumn{2}{c}{V0} &\multicolumn{2}{c}{V1}& \multicolumn{2}{c}{V2}\\
\hline\noalign{\smallskip}
$\fesi$ & &\multicolumn{2}{c}{1.73$\pm$1.55} &\multicolumn{2}{c}{10.68$\pm$1.55}& \multicolumn{2}{c}{1.00$\pm$1.55}\\
$\mgsi$ & &\multicolumn{2}{c}{1.44$\pm$0.91} &\multicolumn{2}{c}{1.02$\pm$0.91}& \multicolumn{2}{c}{1.14$\pm$0.91}\\

Na$_2$O [wt\%] & &\multicolumn{2}{c}{0.021 } &\multicolumn{2}{c}{0.01}& \multicolumn{2}{c}{0.025 }\\
Al$_2$O$_3$ [wt\%]  & &\multicolumn{2}{c}{0.055} &\multicolumn{2}{c}{0.023}& \multicolumn{2}{c}{0.057}\\
CaO [wt\%] & &\multicolumn{2}{c}{0.021} &\multicolumn{2}{c}{0.01}& \multicolumn{2}{c}{0.021}\\
\hline
\end{tabular} 
\end{center}
\end{table}

\subsubsection{Influence of stellar abundances}
\label{vary}

\citet{dornB} investigated the influence of different bulk abundance constraints on interior estimates. In Figures \ref{fig3} and \ref{fig5}, we similarly show this influence on key interior parameters (\renv, \mice, \rsolid, and \rc) for the median abundance estimate (V0, blue), the iron-rich case (V1, light green), and the iron-poor case (V2, dark green) \reve{(Table \ref{bulk})}. As discussed by \citet{dornB}, the largest effects are seen on \rsolid and \rc: if the planets are iron-rich, the core size is significantly larger, which implies a smaller rocky interior (\rsolid) in order to fit mass. The effect on \renv is apparent in the comparison between the iron-rich case V1 and V0. For an iron-rich planet, the density of the rocky interior is higher. In order to fit the mass, \rsolid is smaller. Consequently, to fit the radius, \renv can be larger. For the iron-rich case (V1), the upper 99\% percentile of \renv is 0.02 (HD~219134~b) and 0.04 (HD~219134~c) larger than for V0. Even if the iron-rich case is in agreement with spectroscopic data, we believe that V1 may represent a limitation in  accuracy rather than the actual planet bulk abundance.

\subsubsection{Influence of data uncertainty}

In Figures \ref{fig3} and \ref{fig5}, we also investigate the improvement in constraining interior parameters assuming the hypothetical case of having double the precision on (1) observed mass (light purple), and (2) mass and radius (dark purple). Significant improvement in constraining interior parameters is only obvious for  HD~219134~b when \reve{both mass and radius} precision is doubled. For HD~219134~c, the increase in both mass and radius uncertainty leads to only moderate improvement of parameter estimates. The different potential to improve interior estimates by reducing data uncertainty for planets b and c stems from the fact, that the uncertainties are much smaller for planet c ($\sigma_{R} = 3.1\%$) compared to b ($\sigma_{R} = 5.4\%$). 
\reve{For the considered planets, improved constraints for interior parameters are dominantly gained by a better precision in radius.} This is expected, since mass-radius curves flatten out at higher masses (Fig. \ref{fig1}).

\subsubsection{Influence of prior on \Zenv}

\begin{figure*}[ht]
\centering
 \includegraphics[width = 1.\textwidth, trim = 3cm 0cm 2.9cm 0cm, clip]{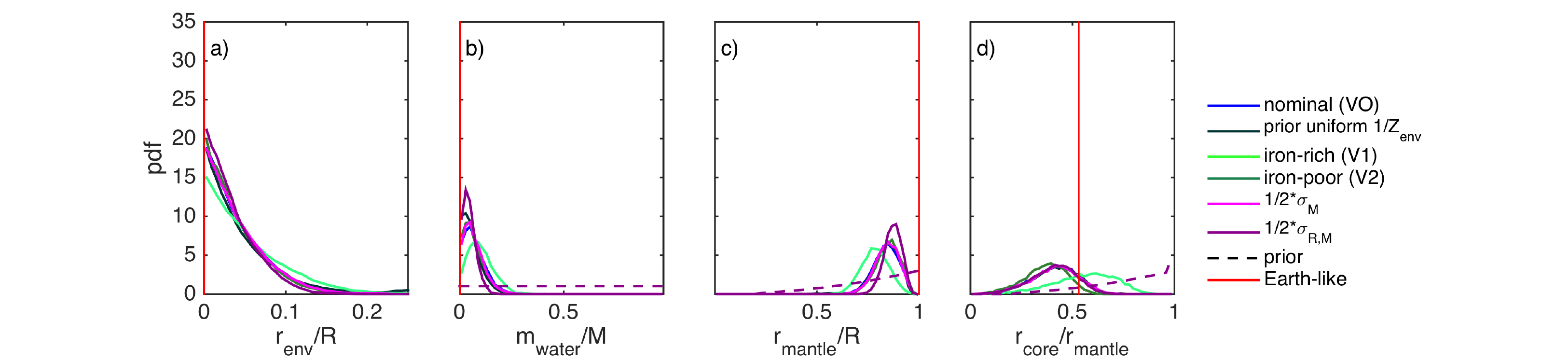}\\
 \caption{Sampled one-dimensional marginal posterior for selected parameters of HD~219134~b: (a) gas radius fraction \renv, (b) water mass fraction \mice$/M$, (c) rock radius fraction \rsolid$/R$, and (d) relative core radius \rc/\rsolid. The posterior distributions depend on precision on bulk abundance constraints (light and dark green curves), and mass and radius uncertainties (light and dark purple curves). For comparison, the Earth-like solution is highlighted in red.}
 \label{fig3}
\end{figure*}

\begin{figure*}[ht]
\centering
 \includegraphics[width = 1.\textwidth, trim = 3cm 0cm 2.9cm 0cm, clip]{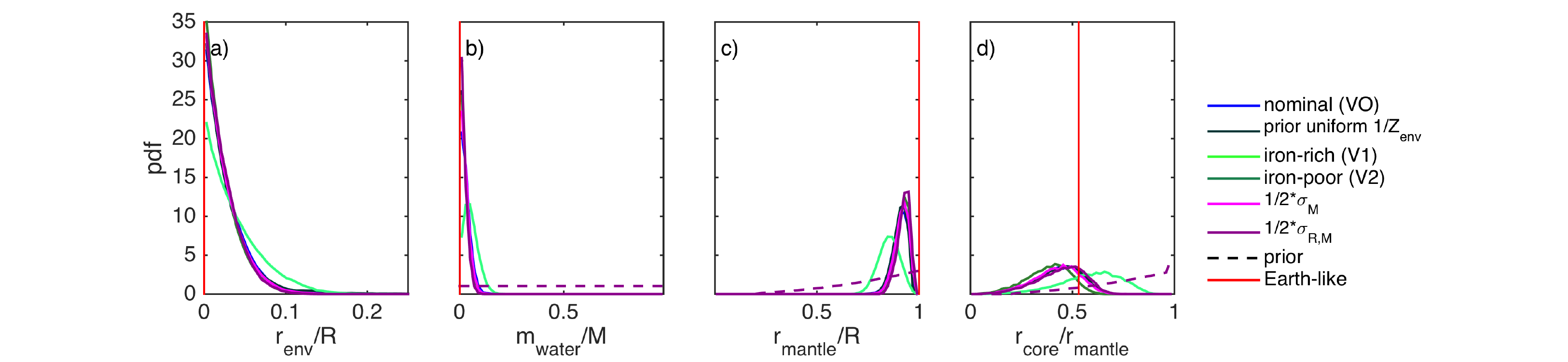}\\
 \caption{Sampled one-dimensional marginal posterior for selected parameters of HD~219134~c: (a) gas radius fraction \renv, (b) water mass fraction \mice$/M$, (c) rock radius fraction \rsolid$/R$, and (d) relative core radius \rc/\rsolid. The posterior distributions depend on precision on bulk abundance constraints (light and dark green curves), and mass and radius uncertainties (light and dark purple curves). For comparison, the Earth-like solution is highlighted in red.}
 \label{fig5}
\end{figure*}

We have shown that the individual parameters \Lenv, \Zenv, and \menv are weakly constrained and thereby are dominated by their prior distributions. However, \renv is well constrained (Figures \ref{fig3} and \ref{fig5}), which is not explicitly a model parameter in this study. Here, we investigate  the effect of different priors on the radius fractions \renv.  An obvious prior to test is on \Zenv. The prior on gas metallicity can be chosen such that it favors \reve{H$_2$-dominated} (uniform in 1/\Zenv) or \reve{enriched} atmospheres (uniform in \Zenv). In Figures \ref{fig3} and \ref{fig5} (comparing blue and dark grey curve), we demonstrate that different priors in \Zenv have only small effects on the possible distribution of radius fractions \renv.

\subsection{Secondary or \reve{primordial} atmosphere?}

The comparison in Figure \ref{fig6} of the inferred \renv (solid lines) to {the threshold thickness} $\Delta R/R$ (dashed areas) shows that the possible atmospheres of planets b and c are significantly smaller than $\Delta R/R$. This indicates that the possible atmospheres {are not dominated by hydrogen} but must be secondary in nature. This provides a simple  {test to identify \reve{H$_2$-rich versus enriched} atmospheres, which may then guide future spectroscopic campaigns to characterise atmospheres (e.g., JWST, E-ELT)}.

\subsection{Comparison to other planets}
\label{comparison}

\begin{figure}[ht]
\centering
 \includegraphics[width = .5\textwidth, trim = 0cm 0cm 0cm 0cm, clip]{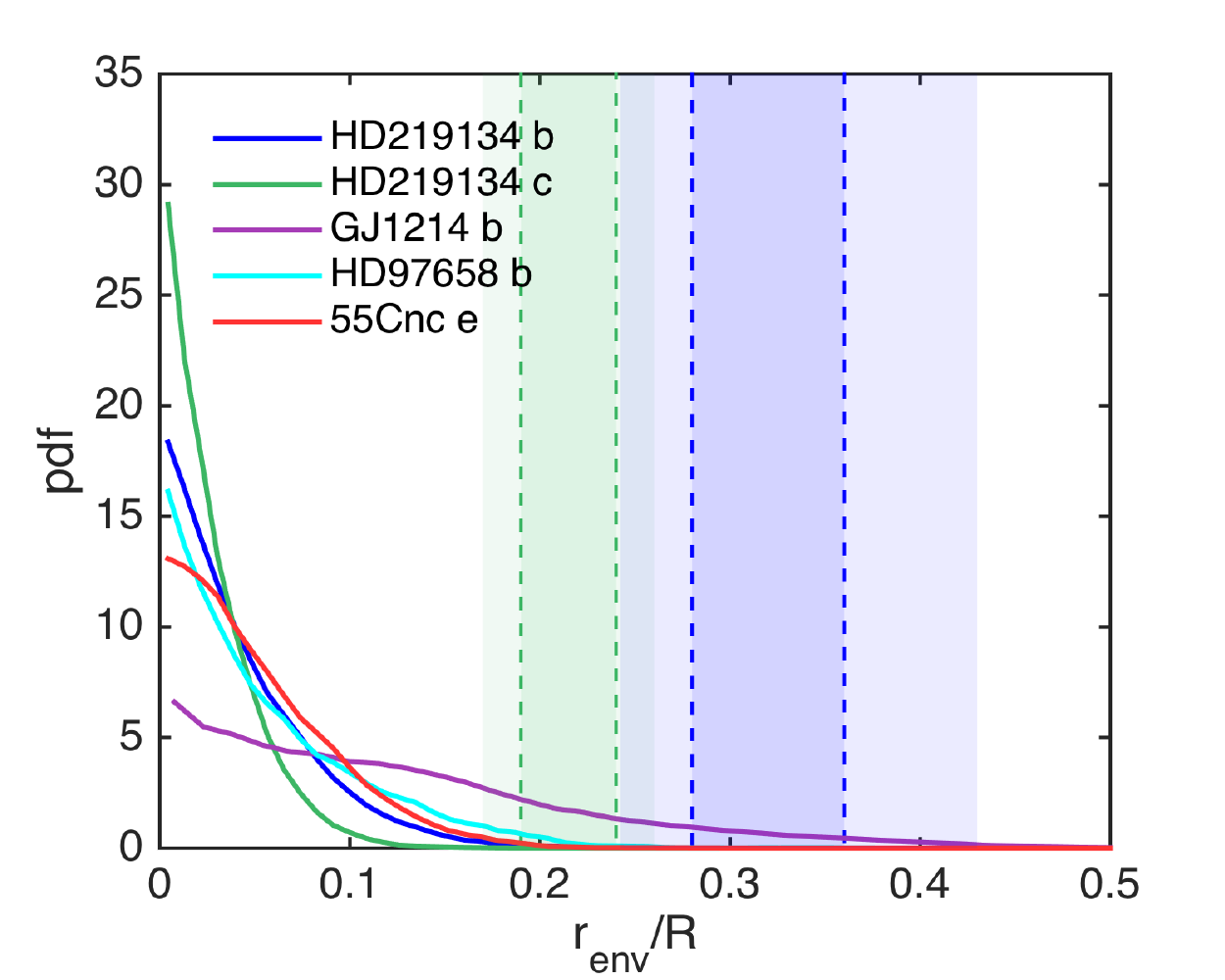}\\
 \caption{Comparison of \renv between five highlighted planets in Figure \ref{fig1}. Inferred radius (solid lines) and approximated {threshold thicknesses $\Delta R/R$ (colored areas with dashed borders)}. $\Delta R/R$ is listed in \reve{Table \ref{deltaR1}} for all five planets.}
 \label{fig6}
\end{figure}

Similarly to HD~219134~b and c, we compare \renv with $\Delta R/R$ (Figure \ref{fig6} and \reve{Table \ref{deltaR1}}) for GJ~1214~b,
HD~97658~b, and 55~Cnc~e. This serves as a benchmark for our proposed determination for \reve{H$_2$-dominated and enriched} atmospheres, since large efforts were put in understanding composition and nature of the atmospheres of the three planets.

For GJ~1214~b, the distribution of \renv is large and overlaps with $\Delta R/R$. The possible atmosphere is consistent with both a \reve{H$_2$-dominated and enriched} atmosphere. Prior to our study, much effort has been invested in characterizing the atmosphere of GJ~1214~b \citep[e.g.,][]{berta, kreidberg}. Studies on interior structure suggested either an hydrogen-rich atmosphere that formed by recent outgassing or a maintained hydrogen-helium atmosphere of \reve{primordial} nature \citep{Rogers2010}. A third scenario of a massive water layer surrounded by a dense water atmosphere has been disfavored by \citet{nettelmann} based on thermal evolution calculations that argued that the water-to-rock ratios would be unreasonable large. Transmission spectroscopy and photometric transit observations revealed that the atmosphere has clouds and/or shows features from a high mean-molecular-mass composition \citep{berta, kreidberg}. 

For HD~97658~b, we find that \renv is very likely smaller than $\Delta R/R$. This suggests an atmosphere that \reve{is enriched and thus possibly} of secondary nature, however, a \reve{primordial} atmosphere cannot be ruled out with certainty. Previous transmission spectroscopy results of \citet{knutson} are in agreement with a flat transmission spectrum, indicating either a cloudy or water-rich atmosphere. The latter scenario would involve photodissociation of water into OH and H at high altitudes. Evidence for this would be  neutral hydrogen escape.  \citet{bourrier} undertook a dedicated Lyman-$\alpha$ line search of three transits but could not find any signature. Any neutral hydrogen escape could happen at low rates only. Consequently, a low hydrogen content in the upper atmosphere is a likely scenario. This is consistent with our findings, that a secondary atmosphere is probable.

For 55~Cnc~e, our prediction clearly indicates a secondary atmosphere, since \renv is significantly lower than $\Delta R/R$. This is in agreement with previous interpretations based on infra-red and optical observations of transits, occultations, and phase curves \reve{\citep{demory12,demory16,angelo}}. This planet has a large day-night-side temperature contrast of about 1300 K and its hottest spot is shifted eastwards from the substellar point \reve{\citep{demory16,angelo}. The implication for the atmosphere is an optically thick atmosphere with inefficient heat redistribution. A bare rocky planet is disfavored \citep{angelo}.} Furthermore, \citet{ehrenreich} give evidence for no extended hydrogen planetary atmospheres \citep[but see][]{tsiaras}. If an atmosphere is present, it would be of secondary nature. Our \reve{approximated approach} leads to the same conclusion.
\reve{Furthermore, the study of 55~Cnc~e's thermal evolution and atmospheric evaporation by \citet{lopez} suggest either a bare rocky planet or a water-rich interior. Although the composition of 55~Cnc~e is a matter of debate, a hydrogen-dominated atmosphere seems unlikely.} 

Also we note that this test holds for Earth and Venus,  {although atmospheric loss mechanisms are very different for them (i.e., Jeans escape and non-thermal escape) \citep{shizgal}.} The threshold thicknesses of possible \reve{primordial} atmospheres are larger than 10 \%, whereas the actual thicknesses are no more than a few percent. Thus, our tests would correctly predict a secondary atmosphere for Earth and Venus.

\begin{table}[ht]
\caption{Threshold {thickness} $\Delta$R/R for different evaporation efficiencies $\eta$. \label{deltaR1}}
\begin{center}
\begin{tabular}{llll}
\hline\noalign{\smallskip}
planet & L$_{\rm x}$ [erg/s] & $\Delta R/R$  &  $\Delta R/R$  \\
&  & ($\eta = 0.01$) & ($\eta = 0.2$)  \\
\noalign{\smallskip}
\hline\noalign{\smallskip}
HD~219134~b& 4$\times 10^{26}$ &0.28&0.36  \\
HD~219134~c &4$\times 10^{26}$  &0.19 &    0.24     \\
GJ~1214~b& 7.4$\times 10^{25}$&0.17    &  0.22  \\
55~Cnc~e & 4$\times 10^{26}$  &0.37  &   0.46   \\
HD~97658~b &1.2$\times 10^{28}$&   0.18 & 0.22\\
Earth &2.24$\times 10^{27}$&  0.12 &0.16\\
Venus &2.24$\times 10^{27}$&  0.15   &  0.21\\
\hline
\end{tabular} 
\end{center}
\end{table}

\begin{table}[ht]
\caption{\rev{Threshold {thickness} $\Delta$R/R for evaporation efficiencies $\eta$ of 0.01 and 95th-percentile of inferred atmosphere thicknesses \renv. $^*$In case of planets for which stellar X-ray luminosities are not available, we assume solar X-ray luminosities. \label{deltaR2}}}
\begin{center}
\begin{tabular}{lll}
\hline\noalign{\smallskip}
planet &95th-percentile of \renv& $\Delta R/R$  \\
\noalign{\smallskip}
\hline\noalign{\smallskip}
Kepler-78 b & 0.15 & 1.0$^*$\\
GJ 1132 b& 0.1 & 0.40$^*$\\
Kepler-93 b& $<$0.05& 0.31$^*$\\
Kepler-10 b& $<$0.05& 0.76$^*$\\
Kepler-36 b& $<$0.05& 0.23$^*$\\
HD 219134 c& 0.13& 0.19\\ %
HD 219134 b& 0.18& 0.28 \\%
CoRoT-7 b &0.1 & 0.59$^*$\\
Kepler-21 b& 0.05& 0.64$^*$\\
Kepler-20 b & 0.05& 0.18$^*$\\
55 Cnc e& 0.18& 0.37 \\%
Kepler-19 b& 0.27& 0.25$^*$\\ 
Kepler-102 e&0.17& 0.10$^*$\\
HD 97658 b& 0.21&0.18 \\ %
Kepler-68 b&0.28 & 0.25$^*$\\ 
Kepler-454 b& 0.28& 0.16$^*$\\
GJ 1214 b& 0.39& 0.17 \\
Kepler-11 d&0.43 & 0.20$^*$\\
Kepler-33 c& 0.48& 0.18$^*$\\
Kepler-79 e& 0.58& 0.25$^*$\\ 
Kepler-36 c& 0.49& 0.30$^*$\\ 
\hline
\end{tabular} 
\end{center}
\end{table}

\begin{figure}[ht]
\centering
 \includegraphics[width = .5\textwidth, trim = 0cm 0cm 0cm 0cm, clip]{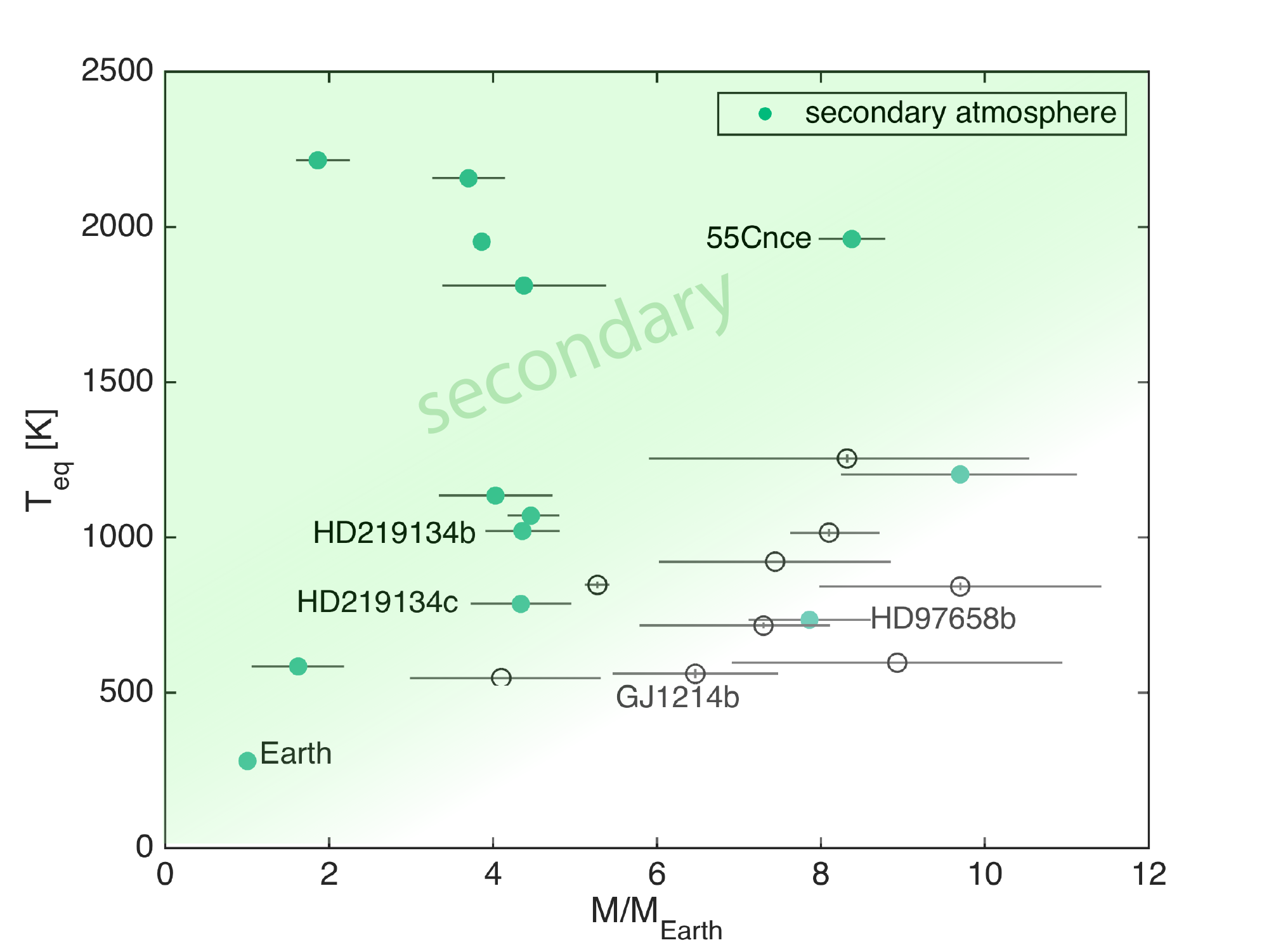}\\
 \caption{Possible origin of atmospheres depending on effective temperature and planet mass. For labeled planets, we use our method described in the text. For unlabelled planets, stellar X-ray luminosities are not available and thus we assume solar X-ray luminosities which is a fair assumption given that the Kepler mission targeted Sun-like stars. Radii and masses of considered planets are shown in Figure \ref{fig1}. }
 \label{fig8}
\end{figure}

A comparison of atmospheric origin on a larger set of exoplanets is limited due to the lack of estimated X-ray stellar luminosities. \reve{For simplicity, we assume solar X-ray luminosities whenever stellar X-ray luminosities are not available, which is a fair assumption given that the Kepler mission targeted Sun-like stars}. Using such simple assumptions, the distribution of planets with secondary atmospheres depends on planet mass and equilibrium temperature to first order (Figure \ref{fig8} and \reve{Table \ref{deltaR2}}). 
In comparison to the tested HD~219134~b and c, most planets have higher equilibrium temperatures and are thus more vulnerable against atmospheric loss. Also, \citet{dornB} concluded that it is unlikely that the planets HD~219134~b, Kepler-10~b, Kepler-93~b, CoRoT-7b, and 55~Cnc~e could retain a hydrogen-dominated atmosphere against evaporative mass loss.

The \reve{possible} transition between secondary and \reve{primordial} atmospheres depending on $T_{\rm eq}$ is  positively correlated with planet mass (Figure \ref{fig8}).  Theoretical photo-evaporation studies \citep[e.g.,][]{JIN2014, lopez2013}  {and the study on observed planets by  \citet{Lecavelier}} predict similar trends, in that planets need to be more massive when receiving higher incident flux in order to retain their \reve{primordial} atmospheres.
For a better understanding of the observed distribution of secondary atmospheres, 
future estimates of X-ray stellar luminosities are required. 

\section{Discussion}
\label{Discussion}

 As already mentioned, the strongest assumption we make is that mass loss is constant over the stellar age.  A more accurate approach is to calculate
\begin{equation}
M_{\rm env,lost} = \int_0^{t_\star} \frac{\pi \eta F_{\rm X} R^2}{E_g} dt,
\end{equation}
where $\eta(t)$ and $F_{\rm X}(t)$ are both functions of time.  The X-ray luminosity evolves over the lifetime of the star, which in turn causes the efficiency of atmospheric escape to evolve.  \rev{Also, the planetary radius $R$ depends on the gas mass fraction that changes over time.} We emphasize that, while the $M_{\rm env,lost} = \dot{M} t_\star$ approximation may lack precision, the logical structure of our approach is robust and accurate.  The reasoning remains that $M_{\rm env,lost}$ (corresponding to a thickness of $\Delta R$) worth of atmosphere may be eroded over the stellar lifetime, so any inferred atmosphere with thicknesses less than this threshold are very unlikely to be \reve{primordial} {(H$_2$-dominated)}. 

\reve{In addition, we assume $T = T_{\rm eq}$ while estimating the threshold thickness (Equation \ref{eq3}). \citet{heng16} finds differences on the order of few tens of percents while approximating the scale height with the isothermal scale height at $T = T_{\rm eq}$. If temperatures are higher, the hydrogen escape would be more efficient and $\Delta$R/R would be higher (and vice versa). The uncertainty on the temperature is accounted for by the variability in $\eta$.}

Furthermore, our estimates of the radius fraction \renv are subject to our choices of interior model and assumptions. Changes in the interior model, especially the atmosphere model, can affect the estimated  \renv as discussed  by \citep{dornA}.  Furthermore, we assume distinct layers of core, mantle, water, and gas. This may not be true as discussed for giant planets \citep{stevenson1985,helled}.  




Following the outlined strategy, it is possible to test for other types of atmospheres (e.g., N$_2$ or CO$_2$-dominated atmospheres). Here, we focussed on an atmosphere type that informs us about  formation processes, i.e. we have assumed that a \reve{primordial} atmosphere is dominated by hydrogen. In principle, a \reve{primordial} atmosphere can be enriched by planetesimal disruption during the accretion. However, initial gas fractions for super-Earths are small and it is not clear whether  atmosphere enrichment can be efficient in these cases nor if metal-enriched thin atmospheres remain well-mixed over long timescales.

We have demonstrated that the possible atmospheres on HD~219134~b and c are very likely to be secondary in nature.  We have shown that this result is robust against different assumptions of bulk abundance constraints and prior choices, as shown for \Zenv. 

Based on bulk density, both planets could be potentially rocky. However, we would expect planets, that are rocky and that formed within the same disk, to roughly lie  on the same mass-radius curve. This is because we expect a compositional correlation, i.e. similar abundances of relative refractory elements \citep[e.g.,][]{sotin07}. The fact, that HD~219134~b and c do not fall on one mass-radius curve, suggests that the larger planet b must harbor a substantial volatile layer. 

\reve{Our use of stellar composition as a proxy for the planet bulk composition excludes Mercury-like rocky interiors. If such interiors were applicable to the HD~219134 planets, the rocky interiors would be iron-rich surrounded by substantially thick volatile envelopes in order to fit mass and radius. It remains an open question whether Mercury-like interiors are common or not.}

\section{Conclusions and Outlook}
\label{Conclusions}

We have presented a method in order to determine the nature of a possible atmosphere. Since close-in planets suffer from evaporative mass loss, the amount of \reve{primordial} atmosphere that can be lost is determined by irradiation from the star, lifetime of the system, and evaporation efficiency. Fortunately, the amount of \reve{primordial} atmosphere loss is weakly dependent on evaporation efficiency and system lifetime \reve{in case of the usually Gyr-old observed exoplanets}. {A comparison between the threshold thickness above which a \reve{primordial} atmosphere can be retained against atmospheric escape and the actual possible atmosphere thickness is a clear indicator of whether an atmosphere is secondary.} We performed this analysis for  HD~219134~b and HD~219134~c.

The possible thicknesses of their atmospheres were inferred by using a generalized Bayesian inference method. For this, we have used the data of planet mass, radius, stellar irradiation, and bulk abundance constraints from the star to  constrain the interiors of HD~219134~b and c. Interior parameters include core size, mantle composition and size, water mass fraction, intrinsic luminosity, gas mass, and gas metallicity. Although individual parameters of the gas layer (\menv, \Lenv, \Zenv) are only weakly constrained, the thickness is well contrained. Inferred thicknesses \renv are robust against different assumed priors and bulk abundance constraints.

We summarize our findings on HD~219134~b and HD~219134~c below:
\begin{itemize}
\item maximum radius fractions of possible gas layers are 0.18 (HD~219134~b) and 0.13 (HD~219134~c),
\item the possible atmospheres are likely secondary in nature,
\item HD~219134~b must contain a significant amount of volatiles.
\end{itemize}

Here, we have proposed a simple quantitative determination of the nature of an exoplanetary atmosphere, that does not include spectroscopic measurement. In order to check our method against planets whose atmospheres are intensively studied, we applied it to GJ~1214~b, HD~97658~b, and 55~Cnc~e. Our predictions agree with previous findings on their atmospheres, and may be tested by future infrared transmission spectroscopy performed on these exoplanets.


\begin{acknowledgements}

We thank Yann Alibert and an anonymous referee for constructive comments.
This work was supported by the Swiss National Foundation under grant 15-144 and PZ00P2\_174028. It was in part carried out within the frame of the National Centre for Competence in Research PlanetS. 

\end{acknowledgements}



\label{lastpage}

\end{document}